\documentclass{ws-ijmpe}
\usepackage[super,compress]{cite}
\usepackage[usenames]{color}
\usepackage{graphicx}
\usepackage{amsmath}
\usepackage{amsfonts}
\usepackage{amssymb}
\usepackage{mathrsfs}
\usepackage{bm}
\usepackage{verbatim}
\usepackage{cancel}

\newcommand{\mlo}{M_{\text{lo}}}
\newcommand{\mhi}{M_{\text{hi}}}

\begin{document}

\markboth{Bingwei Long}{Power counting for nuclear forces in chiral effective field theory}

\title{Power counting for nuclear forces in chiral effective field theory}

\author{Bingwei Long}

\address{Center for Theoretical Physics, Department of Physics, Sichuan University, 29 Wang-Jiang Road\\
Chengdu, Sichuan 610064, China\\
bingwei@scu.edu.cn}

\date{May 8, 2016}
\maketitle


\begin{abstract}
The present note summarizes the discourse on power counting issues of chiral nuclear forces, with an emphasis on renormalization-group invariance. Given its introductory nature, I will lean toward narrating a coherent point of view on the concepts, rather than covering comprehensively the development of chiral nuclear forces in different approaches.
\end{abstract}


\ccode{PACS numbers:21.30.Fe，12.39.Fe}

\section{Introduction}

It has become a widespread trend to build nuclear theories upon the underlying theory of strong-interaction physics---  quantum chromodynamics (QCD). Unfortunately, it is a daunting task to solve QCD directly at low energies where nuclear physics resides. There nevertheless seems to be a working strategy agreed upon by many nuclear physicists. The first step is to have lattice QCD calculate observables of few-nucleon systems, to which one fits the low-energy constants (LECs) of effective field theories (EFTs). EFTs will then be used to deal with larger systems and/or few-nucleon observables that are less amenable to lattice calculations, e.g., scattering observables.

EFTs offer model-independent, controlled approximations for QCD at low energies, and they have several advantages over more phenomenological nuclear models. The most important feature is their ability to make a quantitative statement about theoretical errors, a necessary ingredient to make any quantitative calculation falsifiable~\cite{pra_editoriral}. This ability is attributed to power counting of EFT--- the organization principle that prioritizes among infinitely many contributions that are often expressed in terms of Feynman diagrams. Power counting breaks down all contributions allowed by symmetries into different orders, typically in powers of $Q/\mhi$, where $Q$ generically denotes the external 3-momenta of a process, and $\mhi$ the breakdown mass scale of EFT that is usually tied to the detail of the underlying dynamics. It is clear that EFT works best in cases where separation of scales exists: $Q \ll \mhi$.

Nuclear physics is very interesting in that its rich phenomena indicate a hierarchy composed of multiple length (or mass) scales. Even in the simplest system made up of two nucleons, the $S$-wave scattering lengths are significantly larger than the pion Compton wave length. The $^3S_1$ scattering length--- 5.4 fm--- is the smaller one, but is still three times larger than the pion Compton wave length--- 1.4 fm.

Proliferation of scales makes it possible to develop an EFT for each layer in the hierarchy, potentially simplifying the dynamics by compromising its scope. For instance, the large values of scattering lengths in two-nucleon systems suggests that a good chunk of nuclear physics may be described by the so-called pionless EFT in which pions are integrated out and Lagrangian terms are all contact interactions between the nucleons~\cite{Kaplan_1998tg, vankolck_1stpionless, Bedaque_1997qi}. In systems where a cluster of nucleons are much more tightly bound than the rest, the cluster can be integrated in as new low-energy degrees of freedom, resulting in halo EFT. A good example of which is $N\alpha$ system~\cite{Bertulani_2002sz}.

However, I will go in another direction and will cover the development of chiral EFT (ChEFT) for two-nucleon systems, in which the pion degrees of freedom are explicit. In principle, pionful EFT has a larger kinematic scope, but the inclusion of pions makes it much more challenging to grasp, even in the simplest case of two-nucleon system. The goal is to describe the conceptual development from a maybe slightly idiosyncratic perspective, and I apologize in advance for missing some of the important contributions to the field. While a small portion of the material may be original, the major part is known to the community.

Chiral symmetry of QCD and its explicit and spontaneous breaking are the pillar stones of ChEFT. Not only does it explain the origin of the smallness of the pion mass $m_\pi$, it demands that in chiral Lagrangian the couplings of pions to other particles (including pions themselves) be proportional to the pion momenta or the pion mass~\cite{CCWZ, Weinberg_1968de}. This construction of pion couplings obviously facilitates the EFT expansion in powers of $Q/\mhi$.

In the seminal work of Ref.~\refcite{Weinberg_1978kz}, Weinberg proposed the paradigm of analyzing low-energy QCD processes with chiral Lagrangian. At tree-levels, it necessarily reproduces the well-known PCAC results~\cite{Nambu_pcac, Zhou_guangzhao, Gell-Mann}. More importantly, it was shown that each pion loop brings a suppression factor of $(Q/4\pi f_\pi)^2$, where $4\pi f_\pi \simeq 1.2$ GeV. At the end of the day, every diagram contributing to a process can be counted by a certain power of $Q$, $Q^\nu$, and $\nu$ is always a non-negative number. With these crucial points, it is possible to develop some sort of perturbation theory for low-energy processes involving only Goldstone bosons~\cite{Weinberg_1978kz, Gasser_1983yg, Gasser_1984gg}. Due to its perturbative nature, this theory is often referred to as chiral perturbation theory (ChPT).

It then took a while to bring baryons to the stage because unlike the pions, they are not light; even the nucleon--- the lightest baryon--- has a mass that is comparable to $4\pi f_\pi$. First, a technique was needed to make sure that the baryons always propagate forward so that heavy intermediate states of baryon-anti-baryon pairs are properly integrated out. Second, this technique must in the meantime provide relativistic corrections order-by-order in powers of $Q/m_N$, where $m_N$ is the nucleon mass. Throughout the manuscript I consider only the nucleon, while with some extensions other baryons can be included too. The theoretical tool which does exactly that was developed in Ref.~\refcite{Jenkins_1990jv} by Jenkins and Manohar, called heavy-baryon theory.

Turn to our focus--- nuclear forces. We know phenomenologically that it is necessary to depart from the above perturbative paradigm, for shallow-bound states manifested by a large number of nuclei indicate that nuclear forces are intrinsically nonperturbative. As a theoretical rationale, Weinberg pointed out that the nonperturbative nature of few-nucleon processes stems from the purely baryonic intermediate states, states that are not mixed with the pions~\cite{Weinberg_1991um}. With the appearance of these intermediate states comes small denominators proportional to $(Q^2/m_N)^{-1}$, the so-called infrared enhancement. While this insight helped us understand a priori why nuclear forces are strong, it makes it no longer possible to power count the amplitude of few-nucleon processes in exactly the same fashion as that of one-nucleon processes, due to the presence of $m_N$ in the numerators.

A compromise was proposed by Weinberg. One first organizes the potentials--- two-nucleon irreducible diagrams that are free of pure baryonic states--- into different orders according to standard ChPT counting, and then resums the potentials using the Schr\"odinger equation or the Lippmann-Schwinger equation. It was hoped that the resummed amplitudes somehow inherit the ordering of the potentials. Implementation of this prescription has, however, a much longer history~\cite{Ordonez_1993tn, Ordonez_1995rz, Kaiser_1997mw, Kaiser_1998wa, Epelbaum_1998ka, Epelbaum_1999dj, Entem_2001cg, Entem_2003ft} and continues to be worked upon~\cite{Entem_2014msa, Entem_2015xwa, Epelbaum_2014efa, Epelbaum_2014sza}.

Despite the phenomenological successes, we need to ask for more theoretical satisfactions from any EFT for nuclear forces. After all, the phenomenological nuclear forces~\cite{Machleidt_1987hj, Machleidt_2000ge, Stoks_1993, Stoks_1994wp, Wiringa_1994wb, Gross_2008ps, Perez_2013jpa, Perez_2014yla} had already achieved even better accuracies in describing nucleon-nucleon scattering data.

If not being able to estimate the overall size of two-nucleon amplitudes is merely an academic annoyance, another concern as to the consistency of Weinberg's prescription is more grave: the renormalizability. The $Q$ scaling of a diagram \`a la Weinberg applies under the presumption that the diagram is renormalizable: The ultraviolet regularization scale, say, the cutoff in momentum space $\Lambda$, will be removed by the running of appropriate counterterms once the diagrams at a given order are summed up. Therefore, the absence of $\Lambda$ in the final result justifies the exclusive consideration of $Q$ and ignoring $\Lambda$, when counting the diagram. Conversely, if the final amplitude at certain order shows strong dependence on $\Lambda$, the $Q$ scaling becomes questionable.

The failure of a counting scheme at renormalization test usually suggests that some operators necessary to remove the cutoff dependence are missing at the order under consideration. Stated differently, the proposed power counting underestimated these operators and so they must be promoted. This is how in practice renormalization and power counting impact each other. And it is not terribly surprising. The estimation of LECs in Weinberg's original prescription was based on naive dimensional analysis (NDA). If promotion of a certain LEC is needed on renormalization ground, it simply indicates that the LEC runs with renormalization scale $\mu \sim Q$ so dramatically that its value cannot be reliably estimated by NDA.

In Section~\ref{sec_qcal}, I introduce the general tool, referred to as $Q$-calculus here, by which a Feynman diagram is counted in powers of external momenta. The interplay between renormalization and power counting is then exemplified in Section~\ref{sec_ope}. Some of developments beyond LO are covered in Section~\ref{sec_highorders}, finally a short discussion is offered in Section~\ref{sec_discussions}.

\section{$Q$-calculus}\label{sec_qcal}

In the paradigm proposed by Weinberg~\cite{Weinberg_1978kz}, any Feynman diagram generated by chiral Lagrangian scales in certain powers of $Q$ or $m_\pi$.
I will use a few examples to show how this conclusion is reached.

We will at least need the following few terms of chiral Lagrangian to set up the physics regarding chiral nuclear forces~\cite{Weinberg_1990rz}:
\begin{equation}
\begin{split}
    \mathcal{L} &= N^\dagger \left(i\partial_0 + \frac{\vec{\nabla}^2}{2m_N} \right) N - \frac{\epsilon_{abc}}{4f_\pi^2} N^\dagger \tau_a \pi_b \dot{\pi}_c N - \frac{g_A}{2f\pi} N^\dagger \tau_a \vec{\sigma} N \cdot \vec{\nabla} \pi_a \\
    & \quad - \frac{C_s}{2} \left(N^\dagger N \right)^2 - \frac{C_t}{2} \left(N^\dagger \vec{\sigma} N \right)^2 + \cdots \, ,\label{eqn_lag1}
\end{split}
\end{equation}
where $f_\pi =$ 92.4 MeV is the pion decay constant, $g_A = 1.26$ the nucleon axial charge, and $C_{s, t}$ the $S$-wave nucleon-nucleon contact-interaction couplings. The heavy-baryon formalism has been applied to the nucleon. Operators with more fields, more derivatives, or higher powers in $m_\pi$ are subsumed in the ellipsis. Later in practical calculations, the contact operators are to be given in the more convenient partial-wave basis. The pion-nucleon couplings shown here are all derivative, thanks to chiral symmetry of QCD and its spontaneous breaking.

The analysis proceeds with the presumption that ultraviolet cutoffs will eventually drop out of the investigated diagram. Therefore, when a loop subgraph of the diagram is studied, the loop integration must be over loop momenta around $Q$, that is, we are interested in the contribution of the diagram to long-range physics.

\subsection{Single-nucleon processes}

\begin{figure}
    \centering
    \includegraphics[scale = 0.4]{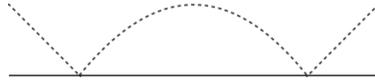}
    \caption{One of the one-loop diagram contributing to pion-nucleon scattering. The solid (dashed) line represents the nucleon (pion).}
    \label{fig:piN}
\end{figure}

I use a one-loop diagram that contributes to pion-nucleon scattering, shown in Fig.~\ref{fig:piN}, as the first example to illustrate the technique~\cite{Jenkins_1990jv, Weinberg_qftbook}. It is assumed that the energies and 3-momenta of incoming and outgoing pions are comparable with $m_\pi$: $Q \sim m_\pi$. In the spirit of studying long-range physics, we need to inspect the integration over the loop momenta around $Q$, $l \sim Q$, where $l$ denotes the 4-momentum of the pion internal line. With that, we can weigh every element in the diagram as follows.

Both vertexes in Fig.~\ref{fig:piN} are the famous Weinberg-Tomozawa term, the second operator in Lagrangian~\eqref{eqn_lag1}, contributing a factor of $Q/f_\pi^2$. The pion propagator scales as $Q^{-2}$. With the nucleon mass subtracted off from the zeroth component of 4-momenta, as conventionally done in heavy-baryon formalism, the nucleon external momentum is of $(Q^2/m_N, Q)$. But the momentum following through the nucleon internal line is of $(Q, Q)$, because the external pion line tends to inject an energy at least of $m_\pi$. So the nucleon internal line is of the size $Q^{-1}$. The integration volume contributes a factor of $Q^4/(16\pi^2)$. The numerical coefficient $16\pi^2$ usually accompanies an integral in which the pion is relativistic, $l_0 \sim |\vec{l}\,|$. Its value should be taken with a grain of salt, for it comes from observations rather than rigorous deductions. In conclusion, the size of the diagram is estimated as
\begin{equation}
    \frac{Q}{f_\pi^2} \left(\frac{Q}{4\pi f_\pi}\right)^2 \, .\label{eqn_piNpc}
\end{equation}

In the loop integral, there may be other momentum modes of $l$ causing one of the two propagators (or both) to be enhanced, compared with the factor of $Q^{-2}$ for the pion and $Q^{-1}$ for the nucleon. However, these modes are so special that they reside only in a small integration volume. For instance, when $l_0$ is near the energy of the incoming pion $k_0$, the nucleon propagator is very close to its mass shell so that it is of the size $(Q^2/m_N)^{-1}$. But this enhancement of the nucleon propagator only happens to a small window of the integration volume $|l_0 - k_0| \sim Q^2/m_N$, so power counting~\eqref{eqn_piNpc} is not altered.

By comparison, special kinematic regions of \emph{external} momenta often call for more cautions. For instance, power counting in Fig.~\ref{fig:piN} needs to be modified when the incoming pion is nonrelativistic in the sense that its 3-momentum is much smaller than the pion mass, $Q \ll m_\pi$. For a recent interesting application of nonrelativistic pion in pion-baryon processes, see Ref.~\refcite{Long_2015pua}.

\subsection{Two-nucleon processes}

\begin{figure}
    \centering
    \includegraphics[scale=0.3]{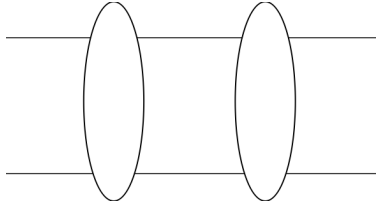}
    \caption{A two-nucleon-reducible diagram. The hallow blob is a subgraph free of two-nucleon intermediate states, i.e., it is a nucleon-nucleon ``potential''.}
    \label{fig:nnloop}
\end{figure}

Processes involving two nonrelativistic nucleons, as is the case in low-energy nuclear physics, present more dramatic changes in $Q$-calculus. Figure~\ref{fig:nnloop} shows a generic loop diagram with two-nucleon intermediate states. Denoting the 4-momentum of one of the nucleon internal lines as $(E/2 + l_0, \vec{l}\,)$, with $E$ being the center-of-mass energy, we can write schematically the loop integral as
\begin{equation}
    \int \frac{d^4 l}{(2\pi)^4} \cdots \frac{i}{\frac{E}{2} + l_0 - \frac{\vec{l}^2}{2m_N}}\frac{i}{\frac{E}{2} - l_0 - \frac{\vec{l}^2}{2m_N}} \cdots
\end{equation}

If $Q$ is again used to represent the external 3-momenta, two nucleons both have 4-momenta of the size $(Q^2/m_N, Q)$. In the narrow shell of phase space $| l_0 - E/2 | \sim Q^2/m_N $, \emph{both} nucleon propagators are enhanced,  combined to give $(Q^2/m_N)^{-2}$. So, the loop integral overall is $\sim m_N Q/(4\pi)$, where $4\pi$ is characteristic of this sort of nonrelativistic integrals.
Referred to by Weinberg as infrared enhancement, the presence of $m_N$ in numerators goes hand in hand with purely nucleonic intermediate states. This analysis gives insights as to why two-nucleon systems are strongly interacting even at low energies, where pion couplings are expected to be weak.

A more mathematically-oriented rationale can be found in Ref.~\refcite{Weinberg_1991um}. The integration over $l_0$ yields (assuming that the potential does not have poles in the $l_0$-plane)
\begin{equation}
    \int \frac{d^3 l}{(2\pi)^3} \cdots \frac{1}{E - \frac{\vec{l}^2}{m_N}}\cdots
\end{equation}
The inclusion of the recoil term $\vec{l}^2/2m_N$ is necessary, or a spurious pole at $E = 0$ will be present. The necessity of including recoil effects in turn suggests the said infrared enhancement.

The recognition of purely nucleonic intermeidate states was conceptually crucial. One immediately identifies two-nucleon irreducible subgraphs as nucleon-nucleon potentials.
The irreducible diagrams still follow standard ChPT counting rule, which is often summarized as follows~\cite{Weinberg_1990rz, Weinberg_1991um, Weinberg_qftbook}: A ChPT diagram is of order $Q^\nu$, and $\nu$ is given by
\begin{equation}
    \nu = 2 - A + 2L + \sum_i V_i \Delta_i\, , \Delta_i = d_i + \frac{f_i}{2} - 2 \, , \label{eqn_nuindex}
\end{equation}
where $A$ is the number of fermion external lines, $L$ the number of loops, $V_i$ the number of vertexes of type $i$. In the definition of factor $\Delta_i$, $d_i$ is the number of derivatives acting on each vertex of type $i$ and $f_i$ the number of fermion lines attached to each vertex.

\begin{figure}
    \centering
    \includegraphics[scale=0.4]{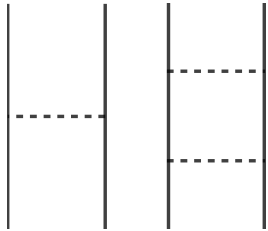}
    \caption{One-pion exchange and its once iteration by two-nucleon intermediate states.}
    \label{fig:ope_iteration}
\end{figure}

It is important to keep in mind that the nucleons coming in and going out of a potential diagram have 4-momenta of the size $(Q^2/m_N, Q)$. This is an element frequently overlooked by beginners. Let us look at how the potential of one-pion exchange (OPE), shown in Fig~\ref{fig:ope_iteration}, is obtained. The nucleon lines of OPE are not always external, because they will be internal lines when OPE is iterated. Their momenta are nevertheless close to shell; therefore, the 4-momentum following through the pion propagator is too of order $(Q^2/m_N, Q)$. Since the pion energy is higher-order, the pion propagator in OPE is taken as instantaneous at leading order (LO):
\begin{equation}
    \frac{-i}{-q_0^2 + \vec{q}^2 + m_\pi^2} = \frac{-i}{\vec{q}^2 + m_\pi^2}\left[1 + \mathcal{O}\left(\frac{Q^2}{m_N} \right) \right] + \cdots
\end{equation}
Not only does this expansion reproduce the on-shell approximation conventionally used in deriving potentials, but it shows that the approximation is under control, i.e., it can be systematically improved in powers of $Q/m_N$.

Figure~\ref{fig:ope_iteration} also shows the once iteration of OPE.
Its size relative to OPE indicates under what circumstance infrared enhancement will force us to iterate OPE to all orders, i.e., to solve the Schr\"odinger or Lippmann-Schwinger equation~\cite{Bedaque_2002mn}:
\begin{align}
    V_\pi &= -\frac{g_A^2}{4f_\pi^2} \bm{\tau}_1 \bm{\cdot} \bm{\tau}_2 \frac{\vec{q}\cdot\vec{\sigma}_1 \vec{q}\cdot\vec{\sigma}_2}{\vec{q}\,^2 + m_\pi^2} \sim \frac{1}{f_\pi^2} \, , \\
    V_\pi G_0 V_\pi &= \frac{1}{f_\pi^2} \frac{m_N Q}{4\pi f_\pi^2} \, .
\end{align}
We can see that nonperturbative treatment of OPE is required when $Q \gtrsim M_{NN} \equiv 4\pi f_\pi^2/m_N$. While $m_\pi^{-1}$ dictates the ranges of pion-exchange forces, $M_{NN}$ characterizes the strength of OPE. It is a little remarkable that low-energy scale $M_{NN}$ emerges through quantum fluctuations, instead of being manifested directly in the Lagrangian.

When $Q \ll M_{NN}$, an EFT based on perturbative OPE can be developed~\cite{Kaplan_1998tg, Kaplan_1998we}. Its phenomenological successes are unfortunately limited~\cite{Fleming_1999ee} because when $Q$ is so low, an even simpler EFT--- pionless EFT--- also applies. It remains nevertheless a useful tool for theoretical explorations. I will focus in the rest of the paper on the case of nonperturbative OPE.

\section{Renormalization of OPE\label{sec_ope}}

Effective field theories are not well-defined until they are regularized by an ultraviolet momentum cutoff $\Lambda$ which excludes the high-energy quantum fluctuations.
The dependence of observables on $\Lambda$ is not wanted, and the procedure to get rid of it by runnings of LECs is called renormalization.

The most obvious objection for sensitivity of observables to $\Lambda$ is that it implies a biased modeling of the underlying physics. Once the form of regularization is chosen, such as dimensional regularization, the Gau\ss ian regulator, etc., it is indeed tempting to fit $\Lambda$ to experimental data, which is equivalent to using $\Lambda$ as an input parameter. This seems reasonable. After all, $\Lambda$ is often linked to the breakdown scale, $\mhi$, where the details of underlying physics becomes important. While it is not wrong to do so, the strategy behind this practice is opposite to that of EFTs. From the perspective of EFTs, the value of $\mhi$ is implied from the size of subleading orders, or more specifically, the values of $g_\nu$, the coefficient in front of the term $(Q/\mhi)^\nu$. Therefore, setting a value for $\Lambda$ and taking that to mean $\mhi$ will effectively assign a value for each of subleading $g_\nu$, with $\nu > 0$. As a result, $g_\nu$'s will be strongly correlated to each other through $\Lambda$. However, in reflecting the unbiased attitude toward the underlying physics we would like $g_\nu$'s to be independent parameters. It is in this sense that the sensitivity to cutoff jeopardizes a model-independent description of short-range physics.

The second reason to remove $\Lambda$ is that $Q$-calculus is based on the assumption that when all the diagrams at a given order are summed up, they will be renormalized. The $\Lambda$ dependence in final results suggests that diagrammatic elements like propagators, vertexes, and loop-integration volumes scale not only with $Q$, but also with $\Lambda$.

In the one-baryon sector, the calculation is perturbative, so at least part of it can be done analytically. Renormalizability is therefore easy to verify. The nonperturbative resummation for few-nucleon processes is amenable only to numerical calculations, and hence the cutoff-independence, or renormalization-group (RG) invariance, cannot be investigated until the calculation is actually done with various values for $\Lambda$. But we would need the power counting, which has yet been verified, to carry out the numerical calculation. To break away from the catch-22 mental picture, one must realize that power counting will necessarily be self-consistent: One proposes a reasonable power counting and then checks its renormalizability, done with the calculations guided by the to-be-verified power counting.

In Weinberg's power counting (WPC)~\cite{Weinberg_1990rz, Weinberg_1991um}, the starting point is NDA, which states that if an operator has $d_\nu$ derivatives (or $m_\pi$) on it, its coefficient $C_\nu$ is suppressed by factor of $\mhi^{-\nu}$. NDA is much more reliable if $\mhi$ is the only scale of the problem, which happens more often in EFTs whose LO is a free theory. However, the emergence of low-energy scale $M_{NN}$ complicates greatly the situation, because one always faces many combinations of $\mhi$ and $M_{NN}$ when dimensionally analyzing $g_\nu$.

I now turn to LO where OPE is resummed to all orders, and discuss how renormalization of this resummation tells us what goes wrong with NDA.

\subsection{Quark-mass dependence of $^1S_0$ counterterms}

The first evidence that NDA does not fulfill RG invariance comes from the quark-mass dependence of $^1S_0$ amplitude. After projecting onto $^1S_0$, LO potential by WPC has two pieces. One is the standard form of a Yuakawa potential with strength proportional to $m_\pi^2$ and a chiral-invariant contact part:
\begin{equation}
    V_{^1S_0}^{(0)}(q) = -\frac{g_A^2}{4 f_\pi^2} \frac{m_\pi^2}{q^2 + m_\pi^2} + C_0 \, ,
\end{equation}
where $\vec{q}$ is the momentum transfer.

Although the Yukawa part is a regular potential by itself, its interference with the contact interaction $C_0$ can produce ultraviolet divergences. Reference~\refcite{Kaplan_1996xu} showed that the divergence is logarithmic
\begin{equation}
    \propto m_\pi^2  \frac{g_A^2 m_N^2}{32\pi^2 f_\pi^2} \ln \Lambda \, .
\end{equation}
This $m_\pi^2$-dependent divergence would need a quark-mass dependent contact term to absorb it, and yet chiral-invariant $C_0$ cannot do that. The need for $m_\pi^2$-dependent counterterms at LO means that they have to be promoted, as compared with NDA. Therefore, $D_2 m_\pi^2$ must be included in the LO $^1S_0$ contact potential, as proposed in Refs~\refcite{Kaplan_1996xu,Beane_2001bc}.

\subsection{Singular attractions of OPE tensor force}

Although conceptually important, the promotion of $D_2 m_\pi^2$ did not have much  phenomenological impact, because chiral extrapolation was not the most urgent application at the time. What caught more attentions was the realization that the tensor part of OPE has renormalization issues too, though a while later, due to the singular attraction of the OPE tensor force.

The singular attraction is most clearly exhibited in coordinate space:
\begin{equation}
    V_T(\vec{r}\,) = \frac{m_\pi^3}{12\pi} \left( \frac{g_A}{2f_\pi} \right)^2 \bm{\tau}_1\bm{\cdot}\bm{\tau}_2\, S_{12}\, \left[1 + \frac{3}{m_\pi r} + \frac{2}{(m_\pi r)^2} \right] \frac{e^{-m_\pi r}}{m_\pi r} \, ,
\end{equation}
where the spin-isospin projector $S_{12}$ is
\begin{equation}
    S_{12} = 3 (\vec{\sigma}_1\cdot\hat{r}) (\vec{\sigma}_2\cdot\hat{r}) - \vec{\sigma}_1 \cdot \vec{\sigma}_2 \, .
\end{equation}
When sandwiched between certain partial waves, $S_{12}$ has negative matrix elements, leading to asymptotic behavior $-1/r^3$ near the origin.

There had been a long history studying singular potentials that are proportional to inverse powers of $r$~\cite{frank}. It was known that the attractive singular potentials do not have a sensible short-range part because it causes a particle to ``fall'' to the center~\cite{landau}. The pathology of their short-range part was reinterpreted in Ref.~\refcite{Beane_2000wh} as the strong sensitivity of observables to the arbitrarily chosen $\Lambda$. A model-independent treatment of short-range physics was then proposed: A short-range counterterm is added and is made to run with $\Lambda$ in such a way that physical observables are independent of $\Lambda$. The point here is that the short-range potential, initially considered as subleading corrections, must appear at LO, together with the singular long-range force. Such an insight concerning importance of the short-range potential is reached a priori, rather than being based on a fit to data.

Reference~\refcite{Nogga_2005hy} applied the same rationale to the OPE tensor force, and concluded that in spin-triplet channels where $V_T(\vec{r}\,)$ is attractive, a counterterm is required on renormalization ground. These channels include $^3S_1 - ^3D_1$, $^3P_0$, and $^3P_2 - {}^3F_2$ in $S$ and $P$ waves. While a constant counterterm in $^3S_1 - ^3D_1$ is already expected by NDA, the need for $P$-wave counterterms at LO is a surprise because they have two powers in momenta (or two derivatives in coordinate space). For instance, $^3P_0$ counterterms after projection are
\begin{equation}
    V_{^3\!P_0} = C_{^3\!P_0}\, p' p + D_{^3\!P_0}\, p' p ({p'}^2 + p^2) + \cdots \, , \label{eqn_V3P0}
\end{equation}
where $\vec{p}$ and $\vec{p}\,'$ are incoming and outgoing center-of-mass momenta, respectively. So the promotion of $C_{^3\!P_0}\, p' p$ to LO is an enhancement of $C_{^3\!P_0}$ by two orders, as compared with NDA.

The immediate criticism to this change of power counting is that infinitely many counterterms will be needed at LO, since with arbitrarily high angular momentum the number of singularly attractive partial waves is infinite. Fortunately, OPE is not always nonperturbative as orbital angular momentum increases for a fixed center-of-mass energy. At some point, OPE becomes subject to perturbation theory and by then renormalization will agree with NDA. A study on where OPE becomes perturbative can be found in Ref.~\refcite{Birse_2005um}.

\section{Higher orders\label{sec_highorders}}

Even though renormalization can veto a power counting scheme, fulfilling RG invariance does not make it absolutely correct. It is not difficult to imagine that there exist many RG-invariant power counting schemes. Therefore, a strategy is needed to find the most efficient one that describes data well with a minimum number of LECs, while avoiding too many fine-tunings.

In the series of works I was involved~\cite{Long_2011qx, Long_2011xw, Long_2012ve}, the strategy is the following:
\begin{enumerate}
    \item Assume that the $Q$-calculus regarding the long-range physics, i.e., pion exchange diagrams does not need modification.
    \item Start with NDA for counterterms since it has the fewest short-range parameters at a given order, while assuming no fine-tuning.
    \item Check the cutoff independence, or the lack thereof.
    \item If RG invariance is violated, identify the higher-order counterterm that is needed for renormalization. Promote it.
    \item Go back to (1) for next order.
\end{enumerate}

It is a better practice to treat the higher-order interactions, either multiple-pion exchanges or counterterms, as perturbation on top of LO, because they are not small corrections until corresponding diagrams are successfully renormalized~\cite{Long_2007vp}. The carefulness here stems from the precaution that two procedures involved here may not commute: Renormalization, in which $\Lambda$ is varied, and the partial resummation of a subset of higher-order contributions.

To go beyond the LO of $NN$ scattering, whose power counting is controlled by renormalization of OPE, it is expected that higher-order counterterms are those renormalizing two-pion exchanges (TPEs)~\cite{Kaiser_1997mw, Kaiser_1998wa}. The leading TPE is suppressed by factor of $(Q/\mhi)^2$, according to the counting rule for irreducible pion-exchange diagrams~\eqref{eqn_nuindex}. So it seems reasonable that subleading counterterms in each partial wave are two orders down from LO . While this is the case for the triplet channels, $^1S_0$ presents an interesting complication.

OPE has the form of the Yukawa potential in $^1S_0$ and subsequently the LO residual cutoff dependence can be shown to be of order $(Q/\Lambda) T^{(0)}$, where $T^{(0)}$ is the LO amplitude ~\cite{PavonValderrama_2007nu, Long_2012ve}. Because the cutoff uncertainty is just one source of theoretical errors, it follows that the LO theoretical error is $\mathcal{O}(Q/\mhi)$. The impact of this finding on power counting is that such a large uncertainty cannot wait until order $Q^2$ to be addressed; a short-range interaction instead of TPEs must appear at order $Q$ to reduce the error. This line of thought eventually leads to a hierarchy of $^1S_0$ counterterms that look just like those of contact EFT:
\begin{equation}
    V_{^1\!S_0} = \sum_{n = 0} C_{2n} Q^{2n}\, , \quad C_{2n} = \frac{4\pi}{m_N} \frac{1}{\mhi^{n} \mlo^{n+1}} \, .
\end{equation}
Without considering the impact of residual cutoff dependences, Ref.~\refcite{Valderrama_2009ei} interestingly reached a similar conclusion, with the only difference being that order $Q$ is still vanishing.

In triplet channels, the LO residual cutoff errors are not large enough to give rise to a non-vanishing order $Q$. One just needs to find out the counterterms that renormalize one insertion of TPE into the LO amplitude. References~\refcite{Long_2011xw} summarized the finding as modified NDA, that in attractive triplet channels the enhancement of counterterms over NDA is uniform. For instance, $C_{^3\!P_0}$, $D_{^3\!P_0}$, and so on, in Eq.~\eqref{eqn_V3P0} are all enhanced by factor $(\mhi/Q)^2$. Although with the same idea, Ref.~\refcite{Valderrama_2011mv} proposed in coupled channels a different scheme with more counterterms. It has yet been resolved why such a difference exists.

\section{Discussions\label{sec_discussions}}

In $Q$-calculus, the running of renormalized LECs against $Q$ was not explicitly considered, although this is implicitly accounted for when cutoff-independence is verified. An ambitious approach is to figure out how renormalized counterterms scale with $Q$ before the full calculations are done, by solving the Wilson's version of RG equation~\cite{Barford_2002je, Birse_2005um, Valderrama_2014vra}.

The nonperturbative nature of nuclear forces makes a ChEFT analysis less transparent than in single-baryon systems, for numerical calculations are often inevitable, one way or another. RG invariance provides us with a valuable tool to scrutinize our understanding toward chiral nuclear forces, before we confront the power counting with experimental data.

Are there other apparatuses than RG invariance to probe the consistency of a power counting?

\textit{Note added in proof}: For a different point of view toward power counting and renormalization, see Ref.~\cite{Epelbaum_Gegelia_09}.

\section*{Acknowledgements}
I thank Chieh-Jen ``Jerry'' Yang for the very enjoyable collaboration, and the Institut de Physique Nucl\'eaire d'Orsay for the hospitality when part of the work was carried out there. Over the years, the conversations with Bira van Kolck, Manuel Pavon Valderrama, Mike Birse, Daniel Phillips, and Harald Grie\ss hammer helped me tremendously in understanding the subject. This work was supported in part by the National Natural Science Foundation of China under Grant No. 11375120.

\end{document}